\documentclass[11pt,twoside]{book}

\usepackage{konkolyproc2}

\usepackage{longtable}

\usepackage{amsmath,amssymb}

\usepackage{graphicx}

\usepackage{lscape}

\usepackage{index}

\usepackage{natbib}

\usepackage{bigdelim}

\usepackage{multirow}

\makeindex

\begin{document}

%%%%%%%%%%%%%%%%%%%%%%%%%%%%%%%%%%%%%%%%%%%%%%%%%%%%%%%%%%%%%%%%%%%%%%%%%%

\pagestyle{myheadings}

\setcounter{equation}{0}\setcounter{figure}{0}\setcounter{footnote}{0}

\setcounter{section}{0}\setcounter{table}{0}\setcounter{page}{1}

\markboth{Clementini}{RR~Lyrae stars in the $Gaia$ era}

\title{RR~Lyrae stars in the {\textit{Gaia}} era}

\author{Gisella Clementini$^{1,2}$}

\affil{$^1$INAF-Osservatorio Astronomico di Bologna, via Ranzani 1, 40127, Bologna, Italy\\}

\affil{$^2$Coordination Unit 7, Gaia Data Processing and Analysis Consortium\\}

\begin{abstract}
$Gaia$, the European Space Agency spacecraft successfully launched on 19 
December 2013, entered into nominal science operations on 18 July 2014 
after a few months of commissioning, and has been scanning the sky to a 
faint limit of $G=20.7$~mag since then. $Gaia$ is expected to observe 
more than a hundred thousand RR~Lyrae stars in the Galactic halo and 
bulge (most of which will be new discoveries), and to provide parallax 
measurements with about 10~$\mu$as uncertainty for those brighter than 
$\langle V \rangle \sim$12-13 mag. 

Status and activities of the spacecraft since launch are briefly 
reviewed with emphasis on preliminary results obtained for RR~Lyrae 
stars observed in the Large Magellanic Cloud during the first 28 days 
of science operations spent in Ecliptic Pole scanning mode and in light 
of the first $Gaia$ data release which is scheduled for summer 2016.

\end{abstract}

\section{Introduction}\label{intro}

$Gaia$ is the ESA cornerstone astrometry mission building on 
the heritage of $Hipparcos$ \citep{perry09}. It is an unbiased all-sky 
($\sim$40,000 deg$^2$) survey that will enable science with one 
billion sources by providing $\mu$as accuracy astrometry (parallaxes, 
positions and proper motions) and  milli-mag optical spectrophotometry 
(luminosities and astrophysical parameters) for sources down to a 
limiting magnitude $G\footnote{$G$ denotes $Gaia$ broad-band white-light 
magnitude.}$ = 20.7 mag, as well as spectroscopy (radial velocities and 
chemistry) for objects brighter than $G$ = 15.3-16.2 mag (and $G >$ 2~mag).

Key-science topics that $Gaia$ will address range from: the study of the 
Milky Way (MW) structure and dynamics to the Galaxy star formation 
history (e.g. \citealt{cigno06}, for a similar study based on 
$Hipparcos$ data),  stellar astrophysics to binaries and multiple 
stars, brown dwarfs and planetary systems to solar system objects, 
galaxies to quasars and the reference frame, and fundamental 
physics to general relativity. 

The backbones of $Gaia$'s science are also i) the discovery of thousands 
new variable sources thanks to repeatedly monitoring the whole celestial 
sphere and, most importantly, ii) the absolute calibration of fundamental 
standard candles of the cosmic distance ladder such as hundreds/thousands of 
RR~Lyrae stars and Cepheids that will have their parallax (hence 
distance) measured by $Gaia$ at $\sim 10$~$\mu$as accuracy. 

A better knowledge of the cosmic distance ladder has a profound 
impact in areas ranging from stellar astrophysics to the cosmological 
model. With the successful launch of $Gaia$ and the release of first 
astrometric data scheduled for mid-2016, this topic has now become 
extremely hot and timely.

\section{{\textit{Gaia}}'s payload and instruments}\label{payload}

$Gaia$ scans the sky from a Lissajous-type orbit around the L2 
Lagrangian point of the Sun and Earth-Moon system, where the spacecraft 
will remain for its 5-year nominal lifetime and possibly for an 
additional year, if the mission is extended. The spacecraft features 
two primary mirrors mounted on a silicon carbide toroidal optical bench, 
each with 1.7$^{\circ} \times 0.6^{\circ}$ field  of view (FoV). The two 
mirrors are separated by a  basic angle of 106.5$^{\circ}$ and share a 
combined focal plane where there are 106 assembled CCDs devoted to different 
functions: 2 wave-front sensor and 2 basic-angle monitor CCDs; 14 Sky Mapper 
CCDs with task of detecting sources entering into $Gaia$'s two fields 
of view; 62 astrometric field CCDs devoted to astrometric measurements 
and providing integrated white-light ($G$-band) photometry over the 
wavelength range: 330-1,050 nm; 7 Blue and 7 Red Photometer (BP and RP) 
CCDs  providing low resolution ($R \sim$20-90) spectrophotometry 
for each source over the wavelength ranges 320-660 and 650-1,000 nm, 
respectively; and 12 CCDs for the  Radial Velocity Spectrograph (RVS) 
obtaining $R=11\,500$ spectra in the Ca triplet (845-872 nm) region for 
sources brighter than $G\sim$15.3-16.2 mag (and $G >$ 2 mag). 
A more detailed description of $Gaia$'s payload, instruments and focal 
plane can be found in \cite{prusti12} and at 
http://www.cosmos.esa.int/web/Gaia/spacecraft-instruments.

The way $Gaia$ scans the sky is due to the spacecraft spinning in 6 
hours around its axis, which points in a direction 45$^{\circ}$ away from 
the Sun, and to the spin axis precessing slowly on the sky (precession 
period of 63 days and 29 revolutions around solar direction in 5 years). 
As a result of the precession, the sky seen by the two fields of view 
every 6 hours changes slowly with time, allowing repeated full sky 
coverage over the mission lifetime. At the same time, due to the 
106.5$^{\circ}$ separation between the two fields of view, objects 
transiting FoV$_1$ at time $t{_0}$ will then transit FoV$_2$ at 
$t=t{_0}$ + 106.5 minutes.  This will then repeat 6 hours later and then 
again 10-30 days later. These figures give rise to an optimum nominal 
scanning law by which over the 5 years $Gaia$ will observe each source 
from 10 to 250 times (the actual number depending on sky position, 
with maximum frequency at $|$${\beta}$$|$ = 45$ \pm 10^{ \circ}$) 
and on average about 70 times in photometry and about 40 times with 
the RVS. The sky coverage of $Gaia$ after 5 years is shown in Fig.~4 of 
\cite{prusti12}. Further details on $Gaia$'s scanning law can be found 
at http://www.cosmos.esa.int/web/Gaia/scanning-law. The  scanning law 
determines the typical cadence of $Gaia$ multi-epoch observations and, 
in turn, bears on the alias patterns we may expect to show up in the 
power spectrum of $Gaia$ time-series data.   

\begin{figure}[!]
\includegraphics[width=0.99\textwidth]{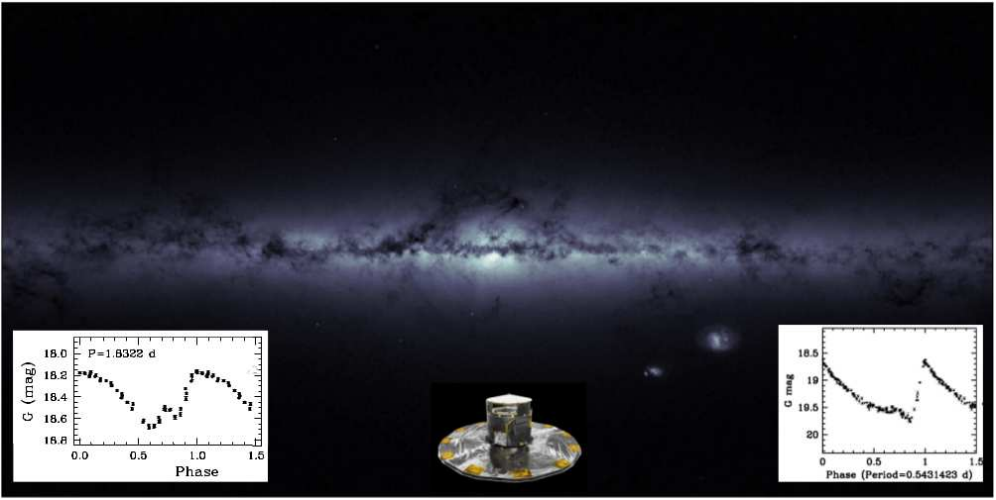}
\caption{Star density map of the sky  produced with $Gaia$'s 
housekeeping data (Credits: E.~Serpell, ESA/Gaia-CC BY-SA 3.0 IGO) showing 
our Galaxy and the two Magellanic Clouds. The two insets show $Gaia$ 
$G$-band light curves for a Cepheid on the left (from 
http://www.cosmos.esa.int/web/Gaia/iow\_20150528) and an RR~Lyrae 
star on the right (from http://www.cosmos.esa.int/web/Gaia/iow\_20150305) 
observed by $Gaia$ in the Large Magellanic Cloud (LMC) during the first  
28 days of science operations in Ecliptic Pole Scanning Law (EPSL).} \label{img:fig1} 
\end{figure}

$Gaia$ does not send images down to ground. Figure~\ref{img:fig1} shows a 
star density map of the sky observed by the spacecraft produced by 
visualizing the number of stars between $G=13$ and 18 mag detected per 
second by $Gaia$'s fields of view. These stars represent only a very small 
fraction of all detected stars and are used by the attitude control system  
of $Gaia$ to ensure the spacecraft's orientation is maintained with the 
desired precision. The Milky Way and the two Magellanic Clouds are easily 
recognized.

\section{Current status and post-launch performances}
After a few months of commissioning, $Gaia$ entered into nominal 
science operations on July 18, 2014. The routine phase started with 
28 days in EPSL, after which the spacecraft went into Nominal Scanning 
Law (NSL). $Gaia$'s EPSL footprint around the South Ecliptic Pole (SEP) 
intercepts a portion of the LMC containing a large number of RR~Lyrae and 
Cepheids which were monitored repeatedly during these first days of 
science operations. Examples of light curves for an RR~Lyrae and a 
Cepheid observed by $Gaia$ in the LMC during the EPSL are shown in the 
two insets of Fig.~\ref{img:fig1}. 

$Gaia$ data processing is handled by the Data Processing and 
Analysis Consortium (DPAC), an ensemble of approximately 450  
scientists and software developers  from 20 different (primarily 
European) countries, organized in 6 Data Processing Centers and 9  
scientific Coordination Units, each having its own specific tasks. 
DPAC processing takes place in a cyclic way  and with continuous 
interaction and exchange among different CUs and $Gaia$'s Main Data 
Base which resides at  the European Space Astronomy Centre (ESAC) 
(see http://www.cosmos.esa.int/web/Gaia/data-processing). Processing 
of the variable sources observed by $Gaia$ is the task of Coordination 
Unit 7 (CU7) whose main Data Processing Center is at ISDC in Geneva.

$Gaia$ is now fully operational, scanning the sky  to a faint 
limit of $G=20.7$~mag (and completeness at $G=20$~mag) since the start of 
science operations and on average collecting data for 50 million stars 
per day. For comparison, faint limit and completeness of $Gaia$'s 
predecessor, $Hipparcos$, are 12 and 7.3-9.0 mag, respectively.    
Over the first year of operations $Gaia$ has collected more than 272 
billion astrometric measurements, 54.4 billion $BP$, $RP$ photometric 
measurements and 5.4 billion RVS spectra
(see http://www.esa.int/Our\_Activities/ Space\_Science/Gaia/Gaia\_s\_first\_year\_of\_scientific\_observations).
Current magnitude limits are:  $2 < G < 20.7$~mag for photometry 
and astrometry, and $2 < G \leq$ 15.3-16.2 mag for the RVS. Stars 
brighter than $G= 3$ mag are imaged by the Sky Mapper CCDs 
\citep{prusti14}. DPAC has started the cyclic processing 
and first tests were made on the EPSL dataset.

Post-launch performances have been derived after conclusion of $Gaia$ 
commissioning and standard errors of the end-of-mission $Gaia$ photometry, 
astrometry and spectroscopy have been re-assessed. Tables and plots showing 
$Gaia$ post-commissioning performances can be found at
http://www.cosmos.esa.int/web/ Gaia/science-performance.

\section{{\textit{Gaia}}'s RR~Lyrae stars}\label{CU7} 

\cite{eyer+00} predict $Gaia$ to  observe about 70,000 RR~Lyrae stars 
in the Galactic halo (based on estimates of the RR~Lyrae density in the 
MW halo by \citealt{sun91}) and an additional 15,000-40,000 RR~Lyrae in 
the MW bulge (based on  MACHO and OGLE detection rates available at the 
time). However, these might be underestimates, as current ongoing surveys 
such as OGLE, LINEAR, CATALINA, PanSTARRS, PTF, are constantly reporting 
new discoveries and increased RR~Lyrae densities. In conclusion, likely 
$Gaia$ will significantly revise upward the census of Galactic RR~Lyrae 
stars both in the MW and in some of its close companions. Over a hundred 
thousand Galactic RR~Lyrae are expected to be observe by $Gaia$, but
compared to $Hipparcos$, only 186 such variables were observed, of which 
only RR~Lyrae itself has an accurate enough parallax 
($\sigma_{\pi}/\pi \sim$18\%). 
End-of-mission, astrometric standard errors, in units of $\mu$as, for 
position, parallax, and proper motion, as a function of $Gaia$ $G$ 
magnitude, for a G2V star with $(V-I)_0$=0.75 mag and $(V-G)_0$=0.16 mag, 
are summarized in Table~1 of \cite{debrui14}. According to these estimates 
all RR~Lyrae brighter than $\langle V \rangle =$12-13~mag will have their 
parallax measured by $Gaia$ to $\sim 10$~$\mu$as, whereas 
individual accuracies will range between 17 to 140 $\mu$as for RR~Lyrae 
stars in Galactic globular clusters with 
horizontal branch luminosity between $V \sim$14 and 18 mag. 

\begin{figure}[!]
\includegraphics[width=1.0\textwidth]{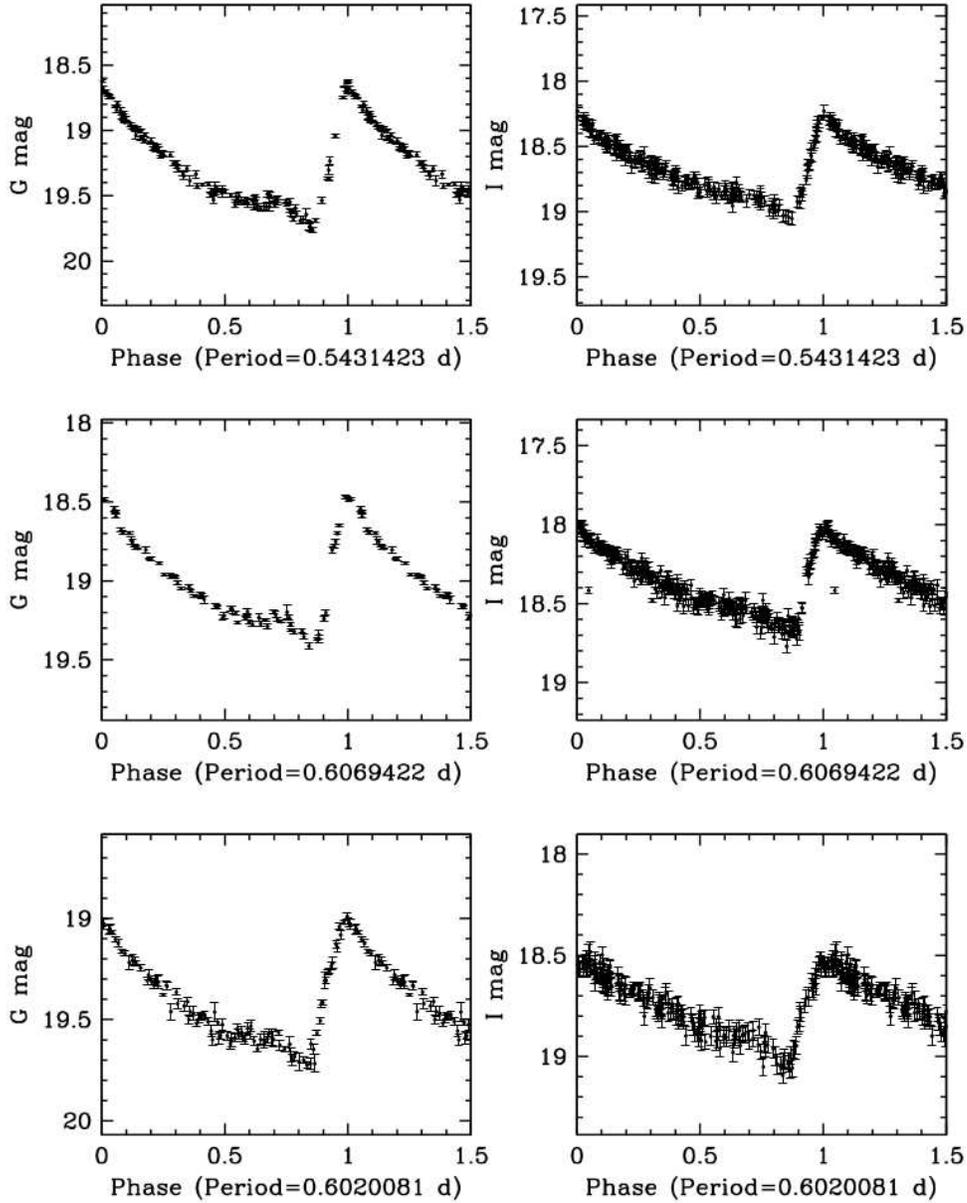}
\caption{Left panels: $G$-band light curves of fundamental-mode RR~Lyrae 
stars in the LMC observed by $Gaia$ during the 28 days of EPSL. 
Right panels: $I$-band light curves obtained for the same stars by the 
OGLE IV survey. 
From http://www.cosmos.esa.int/web/Gaia/iow\_20150305.} 
\label{img:fig2} 
\end{figure}

Processing of the variable sources observed by $Gaia$ is handled by 
CU7, that analyzes the calibrated $G$, $BP$ and $RP$ photometry produced 
by CU5 to identify variable sources. The CU7 processing chain comprises a 
number of different modules and work-packages that perform the variability 
detection, characterize and classify the sources found to vary, and finally 
produce period, amplitude, mean magnitude, modeled light curves, and 
stellar parameters that fully typify the confirmed variables. A specific 
work-package of the CU7 chain is dedicated to the RR~Lyrae stars (and the 
Cepheids) and outputs final attributes for these variable stars, including 
classification in types according to the pulsation mode and detection of 
double-mode pulsation and/or other secondary periodicities.  
In spring 2015 the CU7 pipeline was tested on data collected by $Gaia$ 
during 28 days of EPSL and 3 days of NSL. About 70 million sources were 
received from CU5 and processed by CU7. Over twelve hundred RR~Lyrae stars 
were identified, about half of them are new discoveries. Figure~\ref{img:fig2} 
shows examples of the $G$-band light curve for RR~Lyrae stars in the LMC 
observed by $Gaia$ during EPSL. $Gaia$'s light curves are folded using 
periods taken from the OGLE~IV catalogue of variable stars in the $Gaia$ 
SEP \citep{sos12}. OGLE $I$-band light curves for these stars are also 
shown for comparison. This figure nicely  showcases the excellent quality 
of $Gaia$'s photometry (median uncertainties of the measurements are around 
0.02 mag) at the faint magnitudes of the LMC RR~Lyrae stars (typical 
average apparent magnitudes are $\langle V \rangle \sim 19.5$ mag) and 
after a first data reduction by CU5 and a first analysis by CU7.

$Gaia$ data have no proprietary rights -- they will become public as soon 
as they have been fully processed and properly validated. Publication of 
$Gaia$'s final catalogue is currently planned for 2022, however, there will 
be a number of intermediate data releases, of which the first one is 
foreseen in 2016. This first release will contain positions and $G$-band 
photometry for all-sky single stars, $G$-band time-series photometry and 
characterization by CU7 of the RR~Lyrae and Cepheids observed during the 
EPSL, and parallaxes and proper motions for about 2 million stars in common 
between $Gaia$ and the Tycho-2 catalogue based on the Tycho-Gaia Astrometric 
Solution (TGAS; see \citealt{mik15} and http://www.cosmos.esa.int/web/Gaia/iow\_20150115 for details).  

\subsection{Science with {\textit{Gaia}}'s RR~Lyrae stars} 
Thanks to multi-epoch monitoring of the whole celestial sphere, $Gaia$ will 
discover and measure positions and proper motions of thousands of RR~Lyrae 
stars in the MW and its surroundings down to the spacecraft limiting magnitude 
and will simultaneously determine chemical and dynamical properties for those 
within reach of the RVS (G$\lesssim$16 mag). These RR~Lyrae will trace the 
ancient ($t > $ 10 Gyr) stellar component all the way through from the Galactic 
bulge, to the disk, to the halo and have the potential to unveil streams, 
faint satellites, stellar overdensities, and remnants left over by past 
interactions and accretions of the MW assembling process. As an example of 
$Gaia$'s potential in this area, Table~1 summarizes the estimated number of 
transits per year at the position of classical and ultra-faint Local Group 
galaxies with RR~Lyrae stars within $Gaia$'s reach.

%%%%\vspace*{-3mm}

\begin{table}[!ht] 
\caption{Estimated number of $Gaia$'s transits per year at the
 position of known close by Local Group galaxies}
\smallskip
\begin{center}
\footnotesize 
\begin{tabular}{lllcccccc}
\tableline
\noalign {\smallskip} 
\multicolumn{1}{l}{Name} & \multicolumn{1}{c}{RA} & \multicolumn{1}{c}{Dec} &\multicolumn{1}{c}{D$^{a}$}& \multicolumn{5}{c}{Year of mission}\\ 
  &  ~~(deg) & ~~(deg)  &  (kpc)& 1 & 2$^{b}$ & 3$^{b}$ & 4$^{b}$ & 5$^{b}$\\
\noalign{\smallskip}
\tableline
\noalign{\smallskip}
MW center     &266.41&$-$29.0&$-$&10& 18& 34& 47&~54\\
Canis Major   &108.14&$-$27.66&~7& 22& 49& 72&~98&121\\ 
Segue~1       &~51.76& ~~16.081&23& 10& 20& 29&~50&~59\\
Sagittarius dSph &283.83&$-$30.54&26& 11& 20& 32&~44&~58\\   
Ursa Major~II  &132.88& ~~63.13&32& 19& 51& 73&138&160\\
Segue~2       &~38.416& ~~20.175&35& ~9& 17& 26&~43&~53\\ 
Willman~1      &162.33& ~~51.058&38& 26& 53& 74&~85&~97\\
Bootes~II      &209.5& ~~12.85&42& 10& 30& 56&~77&~84\\  
Coma Berenices&186.74& ~~23.403&44& 15& 33& 42&~49&~61\\
Bootes~III   &209.3& ~~26.8&47& 15& 34& 85&107&114\\
LMC$^{c}$    &~80.893&$-$69.75&51& 14& 28& 47&~63&~81\\ 
SMC          &~13.186&$-$72.82&64& 20& 39& 54&~73&~87\\
Bootes~I     &210.02& ~~14.5&66& ~8& 23& 49&~68&~77\\
Draco        &260.05& ~~57.915&76& 18& 36& 50&~66&~84\\   
Ursa Minor   &227.28& ~~67.222&76& 18& 33& 46&~64&~78\\
Sculptor     &~15.0375&$-$33.709&86& 21& 35& 48&~65&120\\
Sextans      &153.26&~$-$1.614&86& 11& 20& 27&~58&~66\\
\noalign{\smallskip}
\tableline
\noalign{\smallskip}
\end{tabular}  
\end{center}
\vspace*{-2mm}
\scriptsize
$^{a}$ Distances are from \cite{mccon12}\\
$^{b}$ Cumulative numbers\\
$^{c}$ Estimates for the LMC do not include the EPSL transits.
\end{table}

$Gaia$'s complete census of the Galactic RR~Lyrae will definitely increase 
our understanding of the MW structure and formation. However, even more 
dramatic is the impact $Gaia$ will have on the RR~Lyrae (and Cepheid) 
foundations of the cosmic distance ladder. Both the optical 
luminosity-metallicity ($M_{V} -$[Fe/H]) relation and the 
near/mid-infrared period-luminosity ($PLZ$) relation of RR~Lyrae as well 
as the period-luminosity and period-Wesenheit relations of Cepheids need 
an accurate determination of their zero points in order to reliably calibrate 
secondary distance indicators, and be able to probe cosmologically relevant 
distances ($D \ge$ 100~Mpc). At present, an accurate parallax is available 
only for a handful RR~Lyrae stars. $Hipparcos$ measured the parallax for 
more than a hundred RR~Lyrae in the solar neighborhood, but errors are 
larger than 30\%. Only for RR~Lyr itself ($\langle V \rangle \sim $7.8 mag) 
is the $Hipparcos$ error in the parallax smaller than $\sim$18\%. At present, 
the zero points of the RR~Lyrae  relations are anchored on only 5 Galactic 
RR~Lyrae with $HST$ parallax measured by \cite{bene11}. However, $Hipparcos$ 
parallax of RR~Lyr ($\pi_{H}$ = 3.46 $\pm$  0.64 mas, \citealt{vanl07}) 
differs from the $HST$ parallax ($\pi_{HST}$ = 3.77$\pm$ 0.13 mas, 
\citealt{bene11}) by an amount corresponding to a 10\% difference in 
distance, although the two values are not totally inconsistent given the 
large error of $Hipparcos$ estimate. On the other hand, use of \cite{bene11}'s  
parallaxes is not without concern (see, e.g., Section~3.2.2 in \citealt{mura15}). 
$Gaia$ will provide individual distances from parallaxes measured to 
$\sim 10$~$\mu$as for all RR~Lyrae brighter than $V \sim$ 12 mag (see Table~1 in 
\citealt{debrui14}) and RVS metallicity for those brighter than 
$V \lesssim$ 16 mag. Hence, in a few years from now $\sim 10$~$\mu$as accuracy 
parallaxes will become available for 100-150 Galactic RR~Lyrae spanning 
a large enough metallicity range to  measure both the zero point and slope 
of the $M_{V} -$[Fe/H] and $PLZ$ relations directly from these 
parallax-calibrated sources. Accuracies quoted in \cite{debrui14} are 
end-of-mission estimates. However, parallaxes with sub-milliarcsecond 
accuracy for RR~Lyrae brighter than $\langle V\rangle \sim$ 12 mag may 
already be included in the TGAS catalogue published with $Gaia$'s first data 
release in 2016. 

In summary, the unprecedented precision and accuracy of $Gaia$ parallax 
for the local RR~Lyrae (and Cepheids) with allow (i) the absolute 
calibration via parallaxes of these ``primary" standard candles, (ii) a 
test of the metallicity effects through simultaneous abundance measurements, 
(iii) a re-calibration of the ``secondary" distance indicators and (iv) 
to set up a homogeneous distance ladder in and beyond the Local Group, 
and finally producing a total re-assessment of the whole cosmic distance ladder. 
This will in turn significantly improve our knowledge of the Hubble constant 
($H_0$). Furthermore,  by combining $Gaia$'s photometry, parallax, metallicity 
and radial velocity information it will be possible to better constrain the 
physical parameters of RR~Lyrae stars, test  the pulsation models and their 
input physics,  better determine, for instance, the $p$-factor used to convert 
radial to pulsation velocity in Baade-Wesselink studies, better understand 
double-mode pulsation and the Blazhko effect and many more other phenomena 
occurring in RR~Lyrae stars. This will further improve their use as standard 
candles and stellar population tracers. 

Finally, synergy between $Gaia$ and past, ongoing and future surveys 
covering the full wavelength range from near-UV to the mid-IR, such as OGLE, 
EROS, LINEAR, CATALINA, PTF, ASAS, PanSTARRS, LSST, 2MASS, VVV, VMC from the 
ground, and CRRP, SMHASH, CCHPII from space, along with use of upcoming 
multiplex facilities such as WEAVE, MOONS, 4MOST to complement the RVS 
spectroscopy, will allow a further quantum leap in the science achievable 
with RR~Lyrae stars in the $Gaia$ era. 

\vspace*{-2mm}

\section*{Acknowledgments}
This paper extensively uses information publicly available 
at the ESA web pages (http://www.cosmos.esa.int/web/Gaia), whose creation 
and maintenance is gratefully acknowledged. A special thanks goes to the 
DPAC CU7 members at INAF-OACn, to all of the CU7 team, to CU5 for providing 
the processed time-series photometry, and to all DPAC members. Support is 
acknowledged from PRIN-INAF2014, ``EXCALIBUR'S" (P.I. G. Clementini) and from  
the Agenzia Spaziale Italiana (ASI) through grants ASI I/058/10/0 and ASI 
2014-025-R.1.2015.

\end{document}